\documentstyle[psfig]{caosp}
\begin{document}
\def\teff{$T_{\rm eff}$}
\def\lgg{$\log\,{g}$}
\def\vt{$\xi_{\rm t}$}
\def\vsini{$v\cdot \sin i$}
\def\kms{\,km\,s$^{-1}$}
\pubyear{1998}
\volume{27}
\firstpage{258}
\htitle{Discovery of the secondary star of $\kappa$ Cnc}
\hauthor{T. Ryabchikova {\it et al.}}
\title{Discovery of the secondary star of the HgMn binary $\kappa$ Cancri}
\author{T. Ryabchikova \inst{1}  \and O. Kotchoukhov \inst{2}
\and G. Galazutdinov \inst{3} \and F. Musaev \inst{3}
\and S. J. Adelman \inst{4}}
\institute{Institute of Astronomy, RAS, Moscow, Russia
\and Simferopol State University, Simferopol, Crimea, Ukraine
\and Special Astrophysical Observatory, RAS, Nizhnij Arkhys, Russia
\and Department of Physics, The Citadel, Moultrie Street, Charleston,\\
  SC 29409, USA}
%
\maketitle
\begin{abstract}
A careful investigation of a CCD spectrum of the SB1 system $\kappa$ Cnc
in the spectral region 3800~\AA\ -- 8000~\AA\ resulted in the discovery
of the lines of the secondary star.  We then analyzed several short-wavelength
range Reticon spectra obtained at different orbital phases to find additional
radial velocities. The mass ratio is $m_{\rm A}/m_{\rm B}$ = 2.2 $\pm$ 0.1 and,
from binary spectrum-synthesis, the ratio of radii is
$R_{\rm A}/R_{\rm B}\geq$ 2.
\end{abstract}

\section{Introduction}
\label{intr}

$\kappa$ Cnc (= HR 3623 = HD 78316), one of the best known and studied HgMn
stars, is an SB1 spectroscopic binary. Further, a note in the Bright Star
Catalogue (BSC) indicates that $\kappa$ Cnc is a triple occultation system
with the primary having a rather bright companion ($\Delta$m =0.2 mag).
A third star of 7.8 mag in V is at a distance of 0.3 mas from the primary.

During a study of $\kappa$ Cnc's Mn\,{\sc ii} lines we noticed an asymmetry in
the red wings of all Balmer lines on an echelle spectrum obtained at the 1m
telescope of the Special Astrophysical Observatory (SAO).  A careful
reanalysis of all CCD and Reticon spectra available to us resulted in the
discovery of the secondary's lines in them.

\section{Observations and radial velocity measurements.}
\label{Obs}

We used spectra of $\kappa$ Cnc from several observatories, CFHT, DAO, OHP,
CrAO and SAO. Most have S/N ratios greater than 200, and spectral resolutions
between 35000 and 70000. In our initial search for the possible lines of the
secondary, we carefully investigated our SAO spectrum and studied regions
around the most prominent spectral lines including the Ca\,{\sc ii} K line,
Mg\,{\sc ii} $\lambda$ 4481, and strong Fe\,{\sc i} and Fe\,{\sc ii} lines.
We found a few rather broad, shallow features which all gave the same radial
velocity as that estimated from the Balmer lines.  Then we performed a similar
search in other spectra of $\kappa$ Cnc. Table 1 shows our results.

\begin{table}
\caption{A list of the observations of $\kappa$ Cnc.}
\label{table1}
\small
\begin{center}
\begin{tabular}{lcrrr}
\noalign{\smallskip}
\hline
Midpoint (JD)     & &\multicolumn{2}{c} {Radial Velocity (km s$^{-1}$)}\\
2400000 +         & Phase &   Primary      &   Secondary & Observatory\\
\hline
 44619.078 & 0.197& -15.10 $\pm$ 0.40& & CFHT\\
 44620.955 & 0.490& ~67.87 $\pm$ 0.20& & CFHT\\
 44621.042 & 0.504& ~78.32 $\pm$ 0.50& & CFHT\\
 44979.977 & 0.647& ~77.76 $\pm$ 0.40& -98.0 $\pm$   5.0& CFHT\\
 44981.104 & 0.824& ~28.25 $\pm$ 0.40& ~23.5 $\pm$   5.0& CFHT\\
 48378.695 & 0.824& ~12.48 $\pm$ 0.50& ~42.0 $\pm$   5.0& DAO \\
 48586.697 & 0.798& ~38.58 $\pm$ 0.20& ~-1.5 $\pm$   5.0& OHP \\
 48704.760 & 0.265& ~14.10 $\pm$ 1.00& ~60.0 $\pm$   5.0& DAO \\
 49278.051 & 0.937& -32.00 $\pm$ 1.00& 115.0 $\pm$   5.0& DAO \\
 49472.372 & 0.332& ~33.08 $\pm$ 0.50& & CrAO \\
 50168.062 & 0.149& -44.85 $\pm$ 1.00& 163.0 $\pm$   5.0& DAO \\
 50179.335 & 0.912& -13.60 $\pm$ 1.20& ~98.0 $\pm$   3.0& SAO \\
\hline
\end{tabular}
\end{center}
\end{table}

Figure 1 shows the spectrum of $\kappa$ Cnc near Mg\,{\sc ii} $\lambda$ 4481 at
three different orbital phases. The contribution from the secondary is shown by
a dashed line for each spectrum.

\begin{figure}[hbt]
\centerline{
\psfig{figure=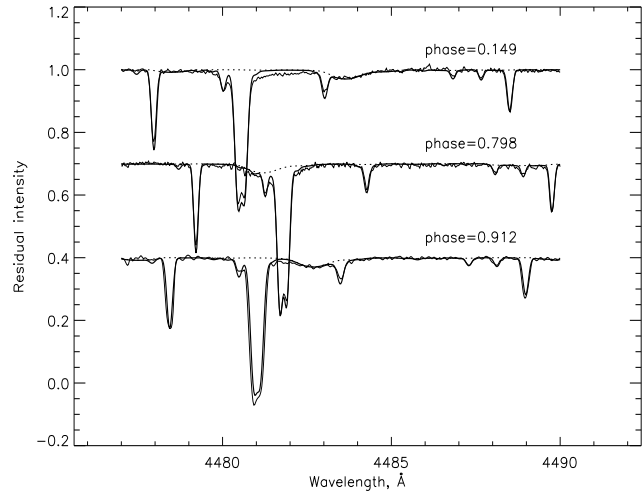,height=7cm,width=9cm}}
\caption{The spectrum of $\kappa$ Cnc at different phases (thin line). The
binary synthetic spectra are shown by thick lines.  The spectra are offset
by 0.3 in residual intensity.}
\label{mg4481}
\end{figure}

\section{Orbital elements and fundamental parameters of the components}
\label{reduct}

Combining our radial velocity measurements with the values from Abt \& Snowden
(1973) and from Aikman (1976), we calculated the orbital elements for
$\kappa$ Cnc using a code by Tokovinin (1992). Due to the small number of the
measurements for the secondary, the orbital parameters for the secondary are
preliminary.

\begin{small}
\begin{tabular}{rlrl}

P(days) : & 6.393190 $\pm$ 0.000013& $ K_{1}$(\kms): & 67.64 $\pm$ 0.58\\
T(JD2400000+): & 40001.936$\pm$ 0.055 &$ K_{2}$(\kms):&148.4 $\pm$ 10.6\\
e:             & 0.137$\pm$ 0.008     & $\gamma$(\kms):&  23.48$\pm$ 0.37\\
$\omega(^\circ)$:&156$^\circ.3 \pm 3^\circ.0$&$m_{A}\sin^3i(m_\odot)$:
& 4.45 $\pm$ 0.76\\    \
&       & $m_{B}\sin^3i(m_\odot)$:&    2.03 $\pm$ 0.20\\
\end{tabular}
\end{small}

\noindent
Within the error limits our orbital elements coincide with those of Aikman.

The effective temperatures and the surface gravities of the components of
$\kappa$ Cnc were obtained by fitting the observed spectrophotometry
(Adelman \& Pyper 1979) and hydrogen line profiles, taking into account our
mass ratio 2.2 $\pm$ 0.1.

Primary: \teff = 13200 K, \lgg = 3.7. Secondary: \teff = 8500 K, \lgg=4.0.\\
$R_{B}/R_{A}$ = 0.48

These atmospheric parameters together with the ratio of radii result in a flux
ratio in the V band of about 11.5, which is equivalent to a magnitude
difference
of $\Delta$m=2.6 mag. It is in good agreement with the $\Delta$m=2.56 mag
difference between the visual magnitude of $\kappa$ Cnc and that of the third
component mentioned in the BSC. The positions of both stars on their
evolutionary tracks indicate that the primary has a mass of about 4.5 $m_\odot$
and is close to the end of its main-sequence life, while the secondary has a
mass about 2 $m_\odot$.

\acknowledgements
We are thankful to Dr. J. Rice and Prof. J. Landstreet who provided us with
spectra of $\kappa$ Cnc obtained at the Canada-France-Hawaii telescope and at
Haute-Provence Observatory. We gratefully acknowledge the use of the {\sc vald}

and {\sc simbad} databases. This work has been partially supported by a Grant
1.4.1.5 of the Russian Federal program ``Astronomy''.


\begin{thebibliography}{}

\article{ Abt, H. A., Snowden, M. S.}{1973}{\apjss}{25}{137}
\article{Adelman, S. J., Pyper, D. M.}{1979}{\aj}{84}{1603}
\bibitem{}
  Aikman, G. C. L.: 1976, Publ. Dominion Astrophys. Obs., {\bf 14}, 379
\bibitem{}
  Tokovinin, A.: 1992, in McAlister H. A, Hartkopf W. I., eds., ASP Conf. Ser.
  Vol. 32, Complementory Approaches to Double and Multiple Star
  Research. Astron. Soc. Pac., San Francisco, p. 573
\end{thebibliography}
\end{document}